\documentstyle[psfig,rotate,12pt,aasms]{article}
\begin{document}
\def\qpr{Q_{pr}}
\def\qprb{Q_{pr,b}}
\def\msun{M_{\odot}}
\def\lsun{L_{\odot}}
\def\etal{{\it et~al.\ }}
\def\eg{{\it e.g.,~}}
\def\ie{{\it i.e.,~}}
\def\de{\partial}
\def\p{{\tilde p}}
\def\R{{\cal R}}
\def\S{{\cal S}}
\def\F{{\cal F}}
\def\ltsima{$\; \buildrel < \over \sim \;$}
\def\simlt{\lower.5ex\hbox{\ltsima}}
\def\gtsima{$\; \buildrel > \over \sim \;$}
\def\simgt{\lower.5ex\hbox{\gtsima}}
\def\ref{\noindent\hangindent.5in\hangafter=1}

\title{THE TURBULENT INTERSTELLAR MEDIUM: GENERALIZING TO A
SCALE-DEPENDENT PHASE CONTINUUM} 
\author{Colin A. Norman}
\affil{Department of Physics and Astronomy, Johns Hopkins University\\
and\\
Space Telescope Science Institute, Baltimore, MD 21218}

\author{Andrea Ferrara}
\affil{Osservatorio Astrofisico di Arcetri,
Largo E. Fermi, 5, 50125 Firenze, Italy}

\begin{abstract}
We discuss the likely sources of turbulence in the ISM and explicitly
calculate the detailed grand source function for the conventional
sources of turbulence from supernovae, superbubbles, stellar winds and
HII regions. We find that the turbulent pumping due to the grand
source function is broad band, consequently expanding the inertial
range of the cascade. We investigate the general properties of the
turbulent spectrum using a simple approach based on a spectral
transfer equation derived from the hydrodynamic Kovasznay
approximation. There are two major findings in this paper: I. The
turbulent pressure calculated from the grand source function is given
by $p_{turb} \sim 10-100 p_{thermal}$.  II. With the scale dependent
energy dissipation from a turbulent cascade the multi-phase medium
concept can be generalized to a more natural continuum description
where density and temperature are functions of scale. Approximate
power-law behavior is seen over a large dynamic range.  We discuss
some of the implications of models for the ISM, particularly
addressing the energy balance of the warm neutral phase. However, this
paper is not concerned with the detailed structure of the ISM of the
Galaxy. Rather, we attempt here to broaden our approach to the global
turbulent structure of the ISM and also to move from the current
many-phase description to a more natural scale-dependent continuum of
phases.
\end{abstract}

\section{INTRODUCTION}

Interstellar turbulence in the multi-phase interstellar medium (ISM)
has an inertial range stretching over at least 9 orders of
magnitude. The substantial viscous and magnetic dissipation at large
wavenumbers requires significant energy sources to maintain the
spectrum. The evidence for a turbulent spectrum in the multi-phase ISM
comes mainly from pulsar observations (Armstrong \etal 1981, Rickett
1990, Narayan 1992, Armstrong \etal 1995) which are consistent with a cascade from scales of
order $\sim 100$~pc to scales down to $10^{11}$~cm. Furthermore, the
power spectrum appears to be consistent with a Kolmogorov law as
reviewed recently by Sridhar and Goldreich (1994). Larson (1981) has
also noted a turbulent spectrum in diffuse neutral clouds and
molecular clouds. Turbulent pressure plays a crucial role in the
various phases of the ISM. This conclusion follows essentially from
the work of Kulkarni \& Fich (1985) for the cold neutral HI component,
and of Reynolds (1985, 1995) for the warm ionized medium. McKee (1990)
argued that turbulent pressure has to represent the dominant pressure
term in the ISM. Similar conclusions, based on the analysis of the
vertical structure of the HI in the Galaxy, were reached independently
by Lockman \& Gehman (1991) and Ferrara (1993); Boulares \& Cox (1990)
also recognized the importance of turbulent pressure. There is general
agreement on the fact that, following the seminal suggestion by
Spitzer (1978), turbulent motion must be associated with energy
deposition fromhigh mass stars in the form of winds, HII regions and
supernova explosions.

There are a number of outstanding problems with the studies of
turbulence in the ISM. It is well known that simple estimates of
turbulent dissipation are too high by three orders of magnitude or so
(Scheuer 1968, Cesarsky 1980, Spangler 1991, Ferriere \etal 1988,
Falgarone \& Puget 1995). The dissipation problem is intimately
related to a Reynolds number that is too low. Using typical numbers
from Spangler we find, in general, a Reynolds number of $\R\sim
400$. Furthermore, the associated dynamic range of the cascade is only
$10^2-10^5$ which is inconsistent with the observations.  The most
recent studies of Shridhar \& Goldreich (1994), Goldreich \& Sridhar
(1995) indicate a solution to the previous problems if the spectrum is
generated by a magnetized four wave cascade in the wavenumber
perpendicular to the field (see also Montgomery \& Mattheus 1995;
Oughton, Priest \& Mattheus 1994; Hossain
\etal 1995).

Let us now outline the structure of the paper and indicate where to
find the principal results. In Section 2 we discuss the general
properties of turbulent cascades and their usefulness in understanding
the ISM. The mathematical formalism for calculating the turbulent
spectrum for a given broad-band source function is presented. We also
derive a formula for the turbulent pressure as a function of the grand
source function. We derive in Section 3 the grand source function for
turbulence in the ISM resulting from supernovae, superbubbles, stellar
winds, and HII regions. The properties of the individual source
functions are given in Table I. Figure 1 shows the individual
contributions to the spectrum and also the total grand source
function. The dynamical equations are then solved numerically with the
grand source function as input and the results are given in Figure~2.
In Section 4 we generalize our analysis to structures called clouds
and using the scale-dependent power input from turbulence we are able
to calculate the multi-phase properties of an ensemble of turbulently
heated clouds. We give a characteristic p-T diagram in Figure 3. In
Figures 4 and 5 we indicate how the density and temperature of clouds
depends on scale. Section 5 discusses the results of these, albeit
idealized, multi-phase calculations and we indicate how they may explain
some of the more commonly observed phenomena in the ISM.

\section{GENERAL PROPERTIES OF TURBULENT ENERGY CASCADES}

In the following we will consider incompressible, homogeneous,
isotropic turbulence. This assumption clearly needs some justification
when discusssing models appropriate to the multi-phase ISM, which is
known to host shock phenomena and highly compressive
processes. However, the bulk of our knowledge on turbulence comes from
terrestrial laboratory experiments and most of them deal with liquids;
in addition one can use those results at least as a reasonable guide
when discussing compressible turbulence. In addition, simulations of
compressible turbulence (Porter \etal 1994) have shown that nonlinear
interactions rapidly transfer most of the energy to noncompressible
modes.

When dealing with turbulence it is normal to introduce a spectral
representation of the velocity field in the fluid. Then, the energy
density in eddies with wavenumber between $k$ and $k+dk$ is given by
$\rho E(k,t)dk$, where $\rho$ is the density of the fluid and $E(k,t)$
is said to be the (tridimensional) spectrum of the turbulence.  Since
turbulence is an intrinsically dissipative process, an external source
is required in order to maintain a steady spectrum. In the inertial
range, however, the dissipation effects can be neglected and the
spectrum approaches an universal form that can be derived form general
principles using dimensional arguments. {\it In absence of an external
pumping}, the spectrum has the well-known expression $E(k,t)\propto
k^{-5/3}$, referred as the Kolmogorov-Obukhoff law.  A general
approach requires to take into account the equation governing the time
evolution of $E(k,t)$, which is called the dynamic equation. In its
integral form the dynamic equation reads (Chandrasekhar 1948,
Heisenberg 1948, Hinze 1975, Lifschitz \& Pitaevskii 1981, Stanisic
1985, Batchelor 1986, Landau \& Lifschitz 1987)
$${\de \over \de t}\int_0^k dk' E(k,t)=\int_0^k dk' F(k,t) -2\nu
\int_0^k dk' k^2 E(k,t)+\int_0^k dk' S(k,t);\eqno(2.1)$$ here $F(k,t)$
is the energy-transfer function, $\nu$ is the kinematic viscosity and
$S(k,t)$ is a source function. The physical meaning of Equation~(2.1) is
transparent and can be understood in the following way. The LHS term
describes the change in the kinetic energy and the dissipation
occurring in large eddies with wavenumber less than $k$. The flow of
energy from large eddies is split in two different contributions: part
of it is transmitted in form of kinetic energy to smaller eddies, and
part is directly dissipated into heat through viscosity. The first two
terms on the RHS describe these two processes, respectively.

A complete solution of the dynamic equation in its general form has
not been found yet; the main difficulty is merely a restatement of the
{\it closure} problem for the hydrodynamic equations. In fact, instead
of being forced to make an assumption to the next-higher-order
velocity correlation, it is necessary here to postulate an explicit
form of $F(k,t)$; this is usually done on the basis of some
physical/dimensional arguments. A general expression for $F(k,t)$ has
been suggested by von Karman (1948) in terms of the interaction
function ${\cal J}(k',k,t)$:
$$F(k,t)=\int_0^\infty dk {\cal J}(k',k,t)=\int_0^k dk' {\cal J}(k',k,t)+
\int_k^\infty dk'' {\cal J}(k'',k,t).\eqno(2.2)$$
The interaction function  describes the contributions to the eddy with
wavenumber $k$ coming from interactions with eddies with $k'<k$ and
$k''>k$; in this sense it has a local character because it does not include
the coupling term coming from the interaction of the two modes $k'$ and
$k''$ themselves. According to von Karman,  ${\cal J}$ can be written as (Hinze 1975, Batchelor 1986, Stanisic 1985)
$$\cases{{\cal J}(k',k,t)=2\alpha E^{3/2-\phi}(k',t)k'^{1/2-\chi}E^\phi
(k,t)k^\chi,& for $k'<k$,~~~~~~(2.3a)\cr
{\cal J}(k,k'',t)=2\alpha E^{3/2-\phi}(k,t)k^{1/2-\chi}E^\phi
(k'',t)k''^\chi,& for $k''>k$,~~~~~(2.3b)\cr}$$
where $\alpha$ is a numerical constant to be determined from experiments,
and the choice of the two exponents $\phi$ and $\chi$ determines the form
of the transfer function. Substitution into Equation~(2.2), and subsequent
integration, give
$$\int_0^k dk F(k,t)=-2\alpha\int_0^k dk' E^{3/2-\phi}(k',t)k'^{1/2-\chi}
\int_k^\infty dk'' E^{\phi}(k'',t)k''^{\chi},\eqno(2.4)$$
Probably the most studied choice corresponds to the set $(\phi,\chi)
=(1/2,-3/2)$ (Heisenberg 1948); this particular form gives origin to the
so-called Heisenberg theory of turbulence, and the second integral in Equation~(2.4)
is usually called eddy viscosity. Unfortunately, this choice brings in a strong
non-local character which contradicts the hypothesis of
statistical independence of the eddies which is at the base of the theory.
For that reason, and because this allows an exact analytical solution
to be found, we have adopted the formulation proposed by Kovasznay (1948),
$$\int_0^k dk' F(k',t)=-\alpha E^{3/2}(k,t)k^{5/2},\eqno(2.5)$$
which, apart from a numerical constant, corresponds to the case
$(\phi,\chi)=(3/2,0)$. This form automatically satisfies the
condition of a local interaction in {\it every} region of the spectrum.
Given the adopted form of the transfer spectrum it follows that Equation~(2.1) can be
more fruitfully written in its differential form
$${\de \over \de t}E(k,t)= -\alpha{\de \over \de k}[E(k,t)^{3/2}k^{5/2}] -2\nu
k^2 E(k,t)+S(k,t).\eqno(2.6)$$
The constant $\alpha$ is of the order of unity; for the Heisenberg
formulation the correct value extrapolated from the experiments is 0.85
(Hinze 1975), which we have used also for the Kovasznay transfer function.

Equation~(2.6) completely describes the evolution of the turbulent spectrum,
provided the appropriate boundary and initial conditions
$$\cases{E(0,t)=E(\infty,t)=0 &  $0\le t < \infty$\cr E(k,0)=E_0(k)&
$0\le k < \infty$}\eqno(2.7)$$
are specified.      

We will concentrate in the following on steady-state solutions of
the dynamic equation. 

\subsection{Steady-state solutions}  

The steady state approximation can be used to analyze the physics
of the turbulent cascade. We begin by deriving the domain of validity
of the steady state assumption. A steady state can be achieved if the
eddy turnover time, $t_e$, is significantly shorter than the interval, $t_i$,
between energy injection events from supernovae, superbubbles, HII
regions and winds.  For a source rate in the Galaxy 
$\bar \gamma_{i}$ and a Galaxy volume $V \sim \pi \varpi^{2} h$ where $h$ and
$\varpi$ are the scale height and the radius of the Galactic disk
respectively, the critical ratio of timescales in the interior of the
bubble radius $R_{0i}$ is

$$ {t_e \over t_i} = {4 \over 3}\left({\bar \gamma_i R_{0i}^{4} \over
\varpi^2 h v_0}\right)\eqno(2.8)$$ 

where $v_0$ is the characteristic turbulent velocity on scale $R_{0i}$.

For supernovae canonical Galactic values (see Table 1) the timescale ratio on
scale $l$ is

$$ {t_e \over t_i} = 20 \left({ l \over R_{0i}}\right)^{ 4 \over 3}.\eqno(2.9)$$

Clearly our steady state assumption is not a very good one close to
the injection scale for supernovae and their may be some time
dependent effects that can influence the cascade which we will not
consider here. However, on scales of order $ l \sim 0.1 R_{0i}$ the
approximation is fair and is increasingly so for larger wavenumbers.
  
We start writing the steady-state dynamic equation in 
nondimensional form:
$${d\Psi_k \over d k} + \R^{-1} k^{1/3}\Psi_k^{2/3}=S(k),\eqno(2.10)$$
where $\R=\alpha v_0/2 \nu k_0$ is proportional to the Reynolds number
corresponding to an dominant eddy of characteristic wavenumber and
turnover velocity $k_0$ and $v_0$, respectively whose actual values
are determined from the input spectrum that we will calculate in this
paper. Note that the eddy turnover time is defined as $t_e=1/(k_0
v_0)$. We have also introduced the auxiliary variable $\Psi_k$, such
that $E(k)=\Psi^{2/3}_k k^{-5/3}$.  For simplicity, we continue to use
for the non dimensional variables the same symbols as for the
dimensional ones.

To proceed further it is necessary to specify the source function
$S(k)$.  We will determine the form of this function for various
assumptions about the energy injection (supernovae, superbubbles,
stellar and galactic winds) in the next Section; here we are
interested in finding a solution of Equation~(2.10) for a model source
function.  If, as we are assuming in this paper, turbulent motions in
the ISM are induced by interacting blast waves, there are some general
constraints on the form of the source function. It is important to
emphasize that we are assuming that the observed turbulent motions and
the associated inferred turbulent pressure in the ISM are ultimately
derived from the kinetic energy of blast waves, winds etc. which, as
discussed later, we assume can act as a source function for a
turbulent cascade with an efficiency of order unity. We also include
reflection from clouds as part of the source function. Bykov \&
Toptygin (1987) have studied in detail the shock-induced turbulence
phenomenon and they find that the turbulent spectrum has, for the
McKee \& Ostriker (1977) model, a $k^{-2}$ dependence in the
short-wavelength regime, while for long wavelengths it is $\propto
k^2$.  Thus, the simplest source function that retains this behavior
is
$$S(k)=\S {k^2 \over 1+k^4}.\eqno(2.11)$$

In the inertial range, $k_0\ll k \ll k_d$, where $k_d$ is the dissipation
wavenumber, the solution closely approaches the constant value 
$$\Psi_k  \simeq {\pi \over 2 \sqrt{2}}\S \simeq \S.\eqno(2.12)$$
It follows from Equation~(2.12) and from the definition of $\Psi_k$ that
$$E\sim \S^{2/3}k^{-5/3}.\eqno(2.13)$$

The energy dissipation rate is given (nondimensionally) by
$$\epsilon_k=\R^{-1} \int_0^k dk k^2 E(k)\simeq 
\R^{-1}\int_1^k dk k^2 E(k)=\R^{-1}\S^{2/3} k^{4/3} \eqno(2.14)$$ 

where the upper bound dominates.

Thus, the dissipation is an increasing function of the wavenumber. 
However, the growth must saturate at a $k$ roughly equal to the
dissipation wavenumber $k_d$, to the limiting value $\epsilon_k\simeq\S$. 
To see this we note that dissipation becomes important when
$$ \R^{-1} k^{1/3}\Psi_k ^{2/3}\sim {\Psi_k \over k},\eqno(2.15)$$
giving the dissipation wavenumber    

$$ k_d\sim \Psi_k^{1/4} \R^{3/4}=\S^{1/4} \R^{3/4}.\eqno(2.16)$$
When we substitute this expression into Equation~(2.14), we find 
$$\epsilon_k\simeq\R^{-1}\S^{2/3} (\S^{1/4} \R^{3/4})^{4/3}= \S.\eqno(2.17)$$ 
Of course this result is also a restatement of the steady-state assumption,
which implies that the rate of energy dissipation must be equal to the energy
injection rate.           

The solution $\Psi_k$
remains nearly constant in the entire inertial range up to the dissipation 
wavenumber $k_d$. For wavenumbers greater than $k_d$ the dynamic equation
is dominated by the dissipative term and can be written as 

$$\R^{-1} k^{1/3}\Psi_k ^{2/3}\simeq \S k^{-2}.\eqno(2.18)$$ 
which has as a solution  
$$\Psi_k=(\R\S)^{3/2} k^{-7/2}.\eqno(2.19)$$
Note that this corresponds to $E(k)\propto k^{-4}$, for the assumed
form of $S(k)$.

If $S(k)$ has a behavior for $k\gg 1$ different from $k^{-2}$, then the
assumption made to obtain Equation~(2.12) is no longer valid. The generic
source function $S(k)\sim \S k^{-n}$, when the same procedure as in Equation~(2.12)
is applied, gives the following solutions $\Psi_k^{(n)}$ depending on 
the value of $n$

$$\Psi_k^{(n< 1)}\sim \S k^{1-n},$$
$$\Psi_k^{(n= 1)}= \S \log k,\eqno(2.20)$$ 
$$\Psi_k^{(n> 1)}\sim \S.$$  

Substituting again into Equation~(2.16) we obtain the expressions for $k_d$:
 
$$k_d^{(n< 1)}\sim \R^{3/(3+n)}\S^{1/(3+n)},$$
$$k_d^{(n= 1)}= \R^{3/4}\S^{1/4}(\log k_d)^{1/4},\eqno(2.21)$$ 
$$k_d^{(n> 1)}\sim \R^{3/4}\S^{1/4}.$$  

As the previous formulae show, the effect of the pumping is to widen the
inertial range beyond the standard value $k_d=\R^{3/4}$ and the gain
is proportional to some power of the source term. This suggests that,
provided that $\S$ is large enough, the inertial range can extend for  
several decades. In addition, if $S(k)$ is not too a steep function of $k$
for $k\gg 1$,
the effect could be very important: for example, for $n=0$, the value
of $k_d$ is $\R^{1/4}\S^{1/3}$ times larger than its standard value. We refer
to this effect as to {\it broad band pumping} of the turbulence. 

An estimate of ${\cal S}$ for the various sources can be obtained
from dimensional analysis once the expansion law is given.
For an expansion law of the source $R \propto t^{\eta}$ the ratio of
the kinetic energy to total energy is, in the thin shell approximation, 

$$ {K \over E_0} = \left( { \eta \over 5\eta - 1}\right).\eqno(2.23)$$ 

It follows, for source i, that ${\cal S}_i$ in dimensionless units is

$$ {\cal S}_i =\left( { \eta \over 5\eta - 1}\right) {\gamma_i R_{0i} 
\over v_0}.\eqno(2.24)$$

For supernovae, using the canonical values given in Tab. 1, 
$$ {\cal S}_{SN} \sim 10^5  \left({\gamma_{SN} \over 0.03~{\rm yr}^{-1}}\right)
\left({R_{0SN} \over 70~{\rm pc}}\right)\left({10 {\rm km~s}^{-1} \over v_0}
\right).\eqno(2.25)$$
\medskip
\centerline{\it 2.2 Turbulent Pressure} 
\medskip

The pressure due to turbulence is given by

$$ p_{turb} = \rho \int_0^\infty dk E(k). \eqno(2.26)$$

Using the assumed expression for the source function $S(k)$ as given by
Equation~(2.11) we integrate Equation~(2.26) and find

$$ p_{turb} = \rho c_s^2\S^{2/3} \int_0^\infty dx~x^{-5/3} \left[ \int_0^x dt
{t^2\over 1+ t^4}\right]^{2/3}, \eqno(2.27)$$

giving 

$$p_{turb} = 6 {\S}^{2/3} p_{th}, \eqno(2.28)$$ 

\section{SOURCE FUNCTIONS OF INTERSTELLAR TURBULENCE} 

There are three catogories for turbulent source functions: (I) shocks
from stellar energy input including supernovae, HII regions, winds and
superbubbles (Spitzer 1982, Ikeuchi \& Spitzer 1984, Bykov \&
Toptygin 1987); (II) Gravitational sources including spiral waves and
bars, tidally driven shocks from interactions, axisymmmetric ($ Q \leq
1$) and non-axisymmetric disk instabilities, turbulent wakes from
orbiting clouds and general shear-fed instabilities in cool spiral
disks (Toomre 1990) and the shear driven by Kelvin-Helmoltz ( Balbus
1988) and Balbus-Hawley (Balbus \& Hawley 1991 , Hawley \& Balbus
1991) instabilities. Note that the Fleck (1982) argument for tapping
the shear of the entire gaseous disk is not valid since it violates
the Rayleigh criterion; (III) Global flows in the ISM including the
global disk-halo circulation (Norman \& Ikeuchi 1989), the Parker
instability on large scales, and the interaction with infalling
high-velocity clouds (Ferrara \& Field 1994) or satellites.

We consider in this paper only the conventional sources in category I
although it is clear that if the shear of the Galaxy could be
efficiently tapped it could dominate the turbulent source functions given here 
and, therefore, the thermal energy balance of the ISM.

In general, for any source type, $i$, the mean number of shocks of Mach
number $\mu$ or greater  crossing a given point in the ISM  (Bykov
\& Toptygin 1987) is

$$ F_i(\mu) = {4 \pi \over 3} \gamma_i {R_i^3\over V}\eqno(3.1)$$

which, for the self similar solutions given in Table 1, can be written as

$$F_i(\mu) = {4 \pi \over 3} {R_{0i}^3\over V} \mu^{-\alpha}\eqno(3.2)$$

where $R_{0i}$ and $\alpha_i$ are also given in Table 1.

The distribution function ${P_{i}}^{(1)}(\mu)$ for primary shocks due
to source, $i$, is

$${P_{i}}^{(1)}(\mu) = - { dF_i \over d\mu} = {4 \pi \over 3}
\alpha_i \gamma_i {R_{0i}^3\over V} \mu^{-(\alpha + 1)} ={\alpha_i P_{0i} \over \mu^{
(\alpha + 1)}}\eqno(3.3)$$

where $P_{0i}=\gamma_i {V_m/V}$, and $V_m={4 \pi / 3} R_{0i}^3$ 
is the volume affected by the energy input. Using the same 
approximation as Bykov \& Toptygin (1987) for the secondary shocks,
\ie shocks are reflected off the clouds considered as dense obstacles
with volume filling factor $f_{cl}$, we find

$${P_{i}}^{(2)}(\mu)={4\pi\alpha_i \gamma_i f_{cl} R_{0i}^3 \over (\mu
- 1)^4V} C(3;\alpha_i), \eqno(3.4)$$

where the coefficient is given by 

$$C(n; \alpha_i) = \int_1^{\sqrt 5} \left({ 5 -x^2 \over 3 x^2 +
1}\right)^{ (\alpha_i + 1)/2} (x - 1)^n dx, \eqno(3.5)$$

and the relevant numerical values for $\alpha_i$ are given in Table 1.

The resultant shock distribution in the ISM including both primaries
and secondaries is

$$ P(\mu) = \sum_i P_{0i} \left[ {1 \over \mu^{\alpha_i + 1}} + { 3
C(3;\alpha_i) f_{cl} \over (\mu - 1)^4}\right]. \eqno(3.6)$$

The spectrum of the velocity field fluctuations is obtained by
assuming a statistically uniform isotropic and stationary ensemble of
shocks of a given Mach number $\mu$. By Fourier analyzing the velocity
field we find a representation for the spectral energy density
distribution, $w_i(k, \mu)$ for source type, $i$, (se Bykov and Toptygin 1987 for details)

$$ w_i = { 4c_s^2 \over \pi} { (\mu - 1)^2 k^2 R_i^3(\mu) \over \left[
1 + k^2R^2(\mu)\right]^2}. \eqno(3.7)$$

The primary spectrum is obtained by averaging over an ensemble of
shocks:

$$S_i^{(1)}(k) =  {\int_1^\infty} w_i(k,
\mu)P_{0i}^{(1)}(\mu)d\mu, \eqno(3.8)$$

which becomes

$$ S_i^{(1)}(k) = c_s^3 \left[ {6 \over \pi}\right] 
\beta Q_i (3+\alpha_i)  f_i(\alpha_i, k^2 R^2_{0i}).\eqno(3.9)$$

We have used the relation
$P_{0i} =\gamma_i (V_m/V) = Q_i/t_m$ (where $Q_i$ is the porosity
factor for the $i$-th source),  and the parameter $\beta = \eta_i R_{0i}
/ c_st_m$, where the maximum radius of the remnant occurs at time $t_m$
(see Cioffi \etal 1988); $\eta_i$  is equal to   $\alpha_i/( 3 + \alpha_i)$. 
The function $f_i$ can be expressed as a combination of hypergeometric functions:

$$f_i(\alpha_i, k^2 R^2_{0i}) =  k^2 R^2_{0i} \sum_{n=0}^2
{ (-1)^n \over (3 - {n \over \delta_i})} F(2, 3 - {n \over \delta_i}, 4 -
{n \over \delta_i}, -k^2 R^2_{0i}). \eqno(3.10)$$

with $\delta_i = (2/3) \alpha_i$.

The grand source function is dependent only on  $\gamma_i$,
$\alpha_i$, and $R_{0i}$. Canonical values are given in Table 1.
Rates have been evaluated using a Salpeter IMF and assuming
that the supernova rate for the Galaxy is one every 30 yr due to stars of
$8M_\odot$ and above. HII regions are assumed to be formed by stars of
$\sim 10 - 20M_\odot$ and above while stellar winds are assumed to be
generated by the supernova progenitors. 

Following Bykov \& Toptygin (1987) we have included one reflection of the primary shocks
from clouds in the following next order expression for the source
function:

$$ S_i(k) = S_i^{(1)}(k)+S_i^{(2)}(k)= c_s^3 \left[ {6 \over \pi}\right]
\beta Q_i (3+\alpha_i) \Big[f_i(\alpha_i, k^2 R^2_{0i}) +$$
$$3 f_{cl}\int_1^{\sqrt 5}d\mu_2
\int_{\mu_2}^{\sqrt 5}d{\mu}_0 \left({ 5 - \mu_0^2 \over 3 \mu_0^2
 +
1}\right)^{(\alpha + 1)/ 2}\left({ \mu_0 - 1 \over \mu_2 -
1}\right)^3 {\mu_00^{ \alpha/3}
 \over \left(\mu_0^{2 \alpha/3} +
k^2 R^2_{0i}\right)}\Big].\eqno(3.11)$$

In evaluating the integral for the reflected shock contribution we
have assumed that the integral is dominated by the pole $(\mu_2 -
1)^{-3}$. The integral over $\mu_0$ is $C(3;\alpha_i)$, and the final
approximation of the integral is:

$$S_i^{(2)}(k) = c_s^3 \left[ {6 \over \pi}\right]
\beta Q_i (3+\alpha_i){3 f_{cl} \over 2}
{C(3; \alpha_i) \over \left[ 1 + k^2R^2_{0i}\right]^2 \left[ \mu_* -
1\right]^2}.\eqno(3.12)$$

>From the papers of Spitzer (1982) and Ikeuchi and Spitzer (1984) we
note that for Mach numbers $\mu \leq 1.5$ soundwaves are radiated with
diminished energy, and for $\mu \leq 2.76$ no reflected shock waves
are found. Conservatively we cut-off $\mu$ at $\mu_* =1.5$. 
A good estimate for the reflected component of the turbulence is then

$$S_i^{(2)}(k) = c_s^3 \left[ {6 \over \pi}\right]
\beta Q_i (3+\alpha_i)6 f_{cl}
{C(3; \alpha_i) \over \left[ 1 + k^2 R^2_{0i}\right)]^2}.\eqno(3.13)$$

The individual source functions $S_i$ are reported in Fig. 1 along
with the grand one. We note that the latter is strongly dominated by
supernovae and therefore the grand source function closely resembles
the simple one used for the analytical solution (Equation~2.10) above: the
approximation $S(k)\propto k^{-2}$ is excellent in the entire inertial
regime. Although the rates ${\gamma_i}$ are similar for the different
sources since they originate from stars of comparable mass, the
$\alpha$ values are different. Since the integrand is a steep function
of $\alpha_i$, the smallest $\alpha_i$ values tend to dominate and for
this reason supernovae and superbubbles the major contributors to the
grand source function. However, the region close to the maximum of
$S_i$ is broader than the one expected from Equation~(2.11). In fact
superbubbles are dominant becuse of their steep $\alpha=9/2$ in
contrast to the $\alpha = 9/7$ of the SNe. Consequently, superbubbles
deposit approximately the same energy as supernovae over a much narrower
range of wavenumbers thus dominating on those scales.

 Secondary shocks contribute sensibly only for small $k$, while their
importance is marginal in the inertial regime. The value of ${\cal S}$
inferred from Fig. 1 is ${\cal S}\sim 50$; then from Equation~(2.28),
$p_{turb}\sim 80 p_{th}$, in agreement with the resulti obtained integrating
numerically on the grand source function. 
In Fig. 2 we give the numerical solution of the
dynamic equation (2.11) using the source function calculated
above. Where a canonical number for the Reynold's number is necessary
we have used a value of $10^5$ using standard values for the viscosity
following Scalo (1987).

\section{TURBULENT THERMAL BALANCE }

In the previous Section we have demonstrated that the dissipation rate
up to a wavenumber $k$ is $\epsilon_k=\R^{-1}\S^{2/3} k^{4/3}$. This
implies that {\it the energy input is scale-dependent}. We now follow
Larson (1981) and generalize our thinking and analysis to a set of
structures and velocites on different scales that forms part of a
general energy cascade. This is not a rigorous step but physically it
is most interesting since it allows us to analyse multi-phase media
with a reasonable scale-dependent energy input.  In other words, we
can study the ISM as a set of structures on different scales
characterized by a specific thermal balance between turbulent heating
and standard cooling. We show that multi-phase equilibria may exist
for a wide range of conditions. For want of a better name we call
these structures clouds but they are, following Larson (1981), clouds,
clumps, subclumps, cores etc.

We first investigate under which conditions turbulent dissipation can
support the various of the ISM. For simplicity, we ignore other
important heating processes such as photoelectric effects from grains
( Wolfire et al 1995) and focus on the structure that emerges with
turbulent dissipative heating. The temperature for a cloud of
wavenumber $T_k$ can be obtained from the energy balance equation,
which is obtained by equating radiative energy losses with turbulent
heating to give

$$ 2 \rho \nu \int_0^k  dk k^2 E(k) = n_e^2 \Lambda(T),\eqno(4.1)$$

where $n_e^2\Lambda(T)$ is the cooling rate per unit volume (see Appendix A), $n_e$
is the electron density, $\rho$ is the mass density of the gas. Because of the
difficulties of understanding the presence of the large inertial range of
the turbulence observed in the ISM, due to the small inertial range 
that is nominally calculated from standard viscosity values 
(Ferriere \etal 1988, Spangler 1991) we have chosen to parametrize 
our collective ignorance in the Reynolds number. Therefore we can 
rewrite Equation~(4.1) as

$$ {2 \rho v_0\over k_0 \R} \int_0^k  dk k^2 E(k) = n_e^2 \Lambda(T).\eqno(4.2)$$

where $k_0$ and $v_0$ are derived form the spectrum.  A further
simplification that allows an equation for the temperature only to be
derived is that the pressure of the ISM is fixed at $\tilde p = nT$.
Multi-phase structure is usually analysed in $p$-$T$ diagrams (Field
1965; Field, Goldsmith \& Habing 1969; Dalgarno \& McCray
1972). However, since in general the net cooling function depends both
on $T$ and $x$, the ionization fraction, at low temperatures the
thermal balance equation must be solved together with the ionization
equation. We will assume that ionization in the ISM is mainly governed
by cosmic-rays (X-ray ionization would give similar results); also, we
will neglect their contribution to the heating with respect to the
turbulent one. The details of the calculation below are given in
Appendix A.

To derive the phase diagrams we now
use Equation~(2.13) as an accurate approximation for $E(k)$; we also use
Equation~(2.16) as an expression for $k_d$. Then Equation~(4.2) can be written 

$$ \tilde p = {T^2 \over \Lambda(T)} \left[ {3\over 2} \left(k_B \S\over 
t_e\right)\left(k\over k_d\right)^{4/3}\right], \eqno(4.3)$$ 
where $t_e$ is defined in Sec. 2.1. For $\S \sim 100$ and $t_e\sim 10^{12}$~s 
then 
$$\tilde p = 2\times 10^{-26} {T^2 \over \Lambda(T)} \left(k\over k_d\right)^{4/3}.
$$

For $k\sim k_d$, the $\tilde p - T$ relation is plotted in Fig. 3. It is clear
then that at the dissipation scale only the very dense, cold phase can absorb
the power and be the dissipative sink. Our conclusion is in contrast to Ferriere
\etal (1988), who attempted to use the less dense warm neutral medium as the
preferential dissipative site. They concluded correctly that the energy 
balance for the warm neutral medium failed by several orders of magnitude. 
However, as we shall show, the dissipation problem is solved by incorporating
scale dependent energy dissipation that follows naturally from the turbulent
cascade. In this case the temperature of the multiphase medium that results
is a function of scale. The density spectrum is also a function of wavenumber.

To see this we fix the thermal pressure at a fiducial value $\tilde p =
3000 \tilde p_{3000}$~cm$^{-3}$~K. Then we directly infer from Equation~(4.3)
that 

$$ \left(k\over k_d\right) = 7.6 \times 10^{21}\left[{\Lambda(T)\over
T^2}\right]^{3/4}.\eqno(4.4)$$

The corresponding $T - k$ relation is shown in Fig. 4. We also derive the
density spectrum directly from Fig. 4 using $n = \tilde p /T$, and this is shown
in Fig. 5. For $\S \simeq 100$ and $\R=10^5$, we have $k_d=\S^{1/4}\R^{3/4}  
\simeq 10^{4} \S_2^{1/4}\R_5^{3/4}$.
The relevant range of interest for the parameters used in this paper is 
$10^{-4} \leq (k/k_d) \leq 1$.

The phase diagrams of our admittedly basic model of the turbulent and
turbulently heated ISM plotted in Figs. 4-5 could represent a
generalization albeit of previous work on multi-phase models of the
ISM. The number of phases required to explain the ISM has been
increasing with time. Our results exhibit a natural extension to a
{\it continuum} of phases in pressure equilibrium, where both the
temperature and density of the ISM are continuous functions of
scale. From an inspection of Figs. 4-5 it is clear that cold, dense
gas is preferentially found on small scales, while the warmer, less
dense material is only at equilibrium on much larger
scales. Obviously, approximate power-laws can be fit to the density
curve and in Fig. 5, over the range of interest, we find $n
\sim (k/k_d)^{.55}$. This shows the approximately fractal nature of
the continuous phase structure in a turbulently heated medium; the
corresponding fractal dimension is $d=2.45$. 

We note the following caveat represented by the possible onset of
thermal instabilities in some temperature ranges (Field 1965; Corbelli
\& Ferrara 1995). This would complicate the thermal structure of the
gas considerably: in particular, some parts of the phase continuum
shown in the phase diagrams may split in two (or more) separate
branches, increasing the overall complexity of the system. The
detailed form of this branching may depend on nonlinear feedback
processes that regulate the pressure of the ISM  and that are
qualitatively introduced and briefly discussed in Sec. 5.

\section{SUMMARY AND CONCLUSIONS} 

We have discussed the physics and mathematical description of the
turbulent energy cascade in the ISM (Section 2). The grand source
function for turbulent energy sources has been derived including the
contribution of conventional sources as supernovae, superbubbles,
stellar winds, and HII regions (Section 3). The results are shown in
Figure 1 and the properties of the individual source functions are
given in Table I. Supernovae dominate over most of the spectral range
although superbubbles can dominate on large scales. We find that the
turbulent pumping due to the grand source function is broad band,
consequently expanding the inertial range of the cascade. The
dynamical equations are solved and the results presented in Figure
2. The turbulent pressure calculated from the grand source function
(Section 2.2) is given by $p_{turb} \sim 10-100 p_{thermal}$. With the
scale dependent energy dissipation from a turbulent cascade the
multi-phase medium concept becomes a continuum description where
density and temperature are functions of scale (Section
4). Approximate fractal behavior is seen over a large dynamic
range. The thermal properties of the ISM on the large scales are
similar to the conventional models. On smaller scales turbulent
heating can dominate and a multi-phase structure can develop for the
hydrodynamic cascade discussed here. A typical p-T diagram is
presented in Figure 3. The derived temperature and density of clouds
in this multi-phase structure are functions of the scale of the
observing measurement and might be regarded as an approximate
fractal. Typical results exhibiting this behaviour are given in
Figures 4 and 5.

We have ignored various potentially important feedback effects in this
paper, regarding them as important to incorporate in the next stage of
calculational complexity. For example, when we use the grand source
function in the calculation of the phase diagram there are temperature
dependences hidden in the $R_{0i}$'s that may create potential feedback
effects on the phase diagram structure shown in Figs. 3, 4, and
5. Such feedback effects will be discussed in a subsequent paper.

This is a steady state calculation with no consideration given to the
initial conditions and how the system can evolve to the steady state.
In fact, it may be difficult for the system to reach this state.  If
the temperature is less than the temperature associated with the
maximum of the cooling curve that a substantial fraction of the gas
must go over, then the hot phase will never be formed. If for other
physical reasons the hot phase is formed then the equilibrium state
can be maintained. The two-equilibrium temperatures are actually two
foci in the phase-plane $T - {\dot T}$ of the time-dependent
equation. The essential point is that the regions of attraction of the
two foci are such that the system will remain attracted to the lower
equilibrium temperature unless it is put over the hump of the cooling
curve and can be attracted to the higher equilibrium temperature. The
condition for this is exactly as described above.

A further complication arises when a magnetized turbulent cascade is
utilized. This is a necessary ingredient since the ISM is certainly
magnetized. The spectrum found is similar to a Kolmogorov spectrum but
is highly anisotropic and is a Kolmogorov spectrum in the wave-number
perpendicular to the magnetic field i.e. $E_M(k_{\perp})\sim
k_{\perp}^{-5/3}$ (Shridhar \& Goldreich 1994, Goldreich \& Shridhar
1995,Oughton, Priest and Mattheus 1994, Montgomery \& Mattheus 1995).
As discussed in detail by Ferriere \etal (1988), Spangler (1991), and
Goldreich \& Shridhar (1995), Alfv\'en waves in the cascade will
propagate without damping in the very highly ionized ISM but will
dissipate rapidly on the edges of clouds consistent with the qualitative
discussion in this paper. However, the details of the dissipative
processes need further study and clarification since there seems to be
a significant discrepancy between the observed inertial range of the
turbulence of up to 10 orders of magnitude and the rather high
dissipation rates inferred for the standard ISM. 

The source of the turbulent pressure that holds up the ISM against its
own weight is clearly evident in our analysis with the principal being
energy input form massive stars via supernova remnants and
superbubbles. The spectrum of the predicted velocity has been
presented and it may be possible to constrain our results by
observations of large scale turbulence via measurements of the large
scale velocity fields (HI, H$\alpha$, IR fine cooling fine structure
lines, recombination lines). As we shall discuss in a later paper,
many other heat sources come into play in the ISM (Wolfire \etal 1995)
including direct shock heating from supernovae and superbubbles,
photoelectric heating from grains, soft X-ray heating etc. (Wolfire
\etal 1995). Many of these heat sources are
scale-dependent. Therefore, the range of validity of turbulent heating
effects for our Galaxy may be limited to the cooler material.

The scale-dependent phase continuum model of the ISM presented here
may be a beginning of a more useful description of the ISM than the current
many-phase models particularly for the understanding and integration of the 
many detailed observations of the ISM.

\acknowledgements

We thank the Aspen Center for Physics, Arcetri Observatory, ESO
Garching, and the Institute of Astronomy, Cambridge whose stimulating
environments, hospitality and support are gratefully acknowledged. We
particularly thank R. Bandiera, T. Heckman, P. Pietrini, Yu. Shchekinov, 
J. Slavin, and M. Spaans for valuable comments.

\bigskip
\centerline{\bf APPENDIX A.}
\medskip

Here we briefly outline the assumptions made in the calculation of
the cooling functions. More details can be found in Ferrara \& Field (1994).
We have adopted the ``on the spot'' approximation in which
diffuse field photons are supposed to be absorbed close to the point
where they have been generated.
The various elements are divided into  ``primary'' ones, which enter the
ionization-thermal equilibrium and ``secondary''ones which do not.  In the
following, H is considered as primary, while other elements (He, C, N,
O, Si, Fe) just contribute to the cooling function.  Secondary
elements, apart from He, are considered to be completely ionized by
the UV field below 13.6 eV; a helium abundance equal to 0.1~H as been
assumed. Double ionization of He has been neglected, and He fractional
ionization has been supposed equal to the H one, $x$. The following
processes have been included in the calculation of the cooling
function $\Lambda(x,T)$: i) free-free emission; ii) H and He
recombination; iii) electron impact ionization of H and He; iv)
electron impact excitation of H and He (n=2,3,4 triplets); v) He
dielectronic recombination; vi) electron and H impact excitation of
secondary elements; excitation of the metastable levels by electron
impact are also included.  The cooling function obtained is shown in
Fig. 2.  by Dalgarno \& McCray (1972) for the range $T\simlt
10^4$~K. At higher temperature the cooling curve has been calculated
using the radiation code by Landini \& Monsignori-Fossi (1990).

The ionization fraction is calculated as follows.
Assume that the energy density in cosmic rays is proportional to the thermal
pressure in the Galaxy. Normalizing to canonical values for the Galaxy we find
the ionization rate 

$$\xi = n_{cr}\sigma c 
\simeq 3\times 10^{-16} \p_{3000}, \eqno(A1)$$
where $\sigma$ is the low-energy cosmic-ray cross section, $c$ is the
speed of light.
We can write the ionization equation as
$$n^2 x^2\alpha(T)=(1-x)n\xi,\eqno(A2)$$
and using Equation~(A1) we find
$$x^2{\alpha(T)\over T}=(1-x)\xi_0,\eqno(A3)$$
with $\xi_0=3\times 10^{-19}$~s$^{-1}$, and $\alpha(T)$ is the radiative
recombination coefficient; Equation~(A3) gives the dependence of $x$ on $T$.

\bigskip
\centerline{\bf APPENDIX B.}
\medskip

In Table 1 we have defined the following symbols and relations:

$$R_{SN}= 14.0 { E_{51}^ {2 \over 7} \over n_0^{ 3 \over 7} \zeta_m^{
1\over 7}} ~~{\rm pc},$$

$$\tau_{SN} = 3.61 \times 10^4  {E_{51}^ {3 \over 14} \over 
n_0^{ 4 \over 7} \zeta_m^{ 5\over 14}} ~~{\rm yr},$$

where $E_{51}$ is the energy of a supernova in units of $10^{51}$ erg,
$\zeta_m$ and $n_0$ are the metallicity and the average density of the
gas respectively (Cioffi \etal 1988).
$R_S$ is the Stromgren radius of the star forming the HII region under
consideration. The sound speed in the neutral and ionized media are
given by $C_{I}= c_s$ and $C_{II}$, respectively.  
$L$ and $L_{SB}$ are the luminosities of  winds and superbubbles, respectively.

\pagestyle{empty}

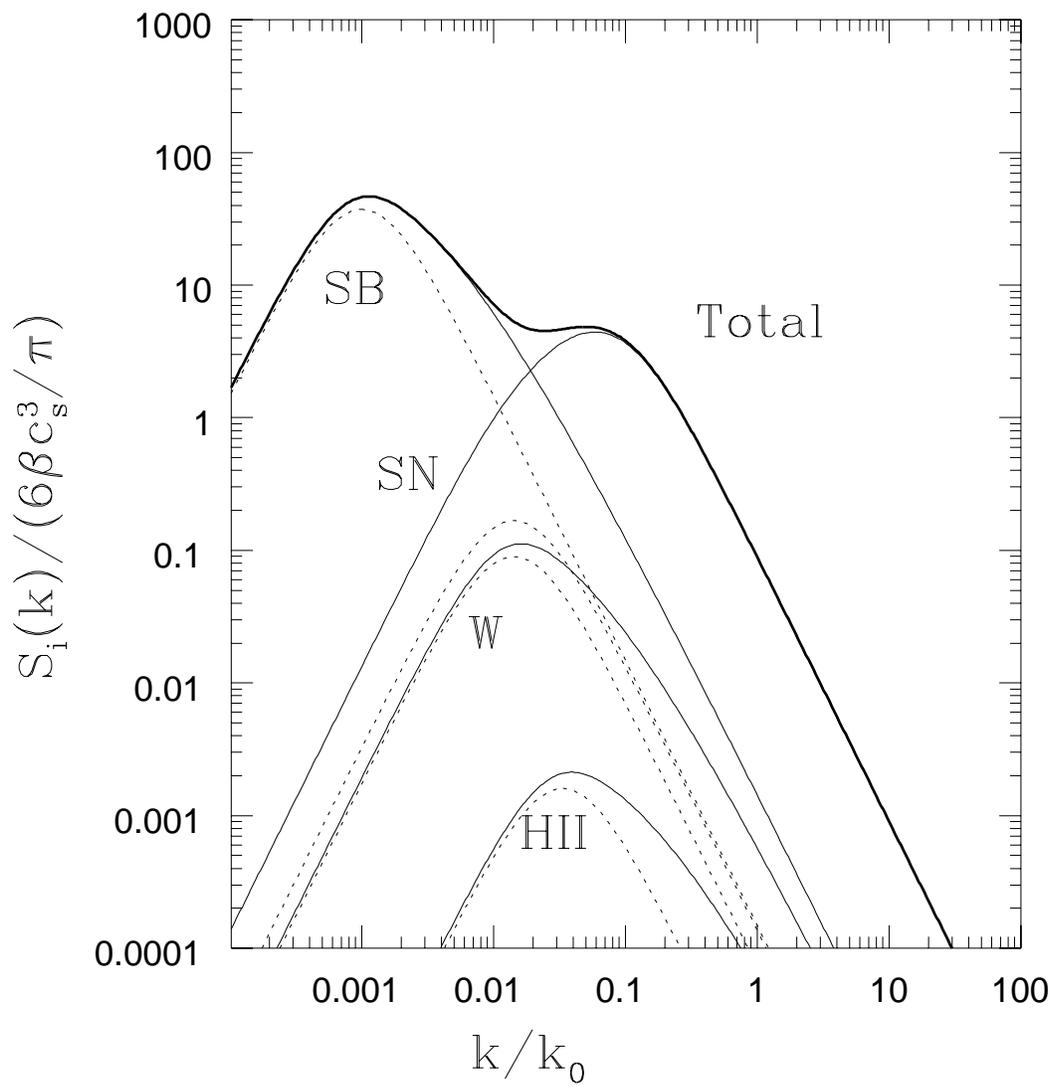
\begin{figure}
\caption{Normalized turbulent source functions $S_i(k)$ for supernovae, 
superbubbles, winds and HII regions. {\it Solid lines} show the sum of primary 
and secondary shock contributions for each source; {\it dashed lines} show 
secondary shocks only. The thick line is the total grand source
function.}
\end{figure}

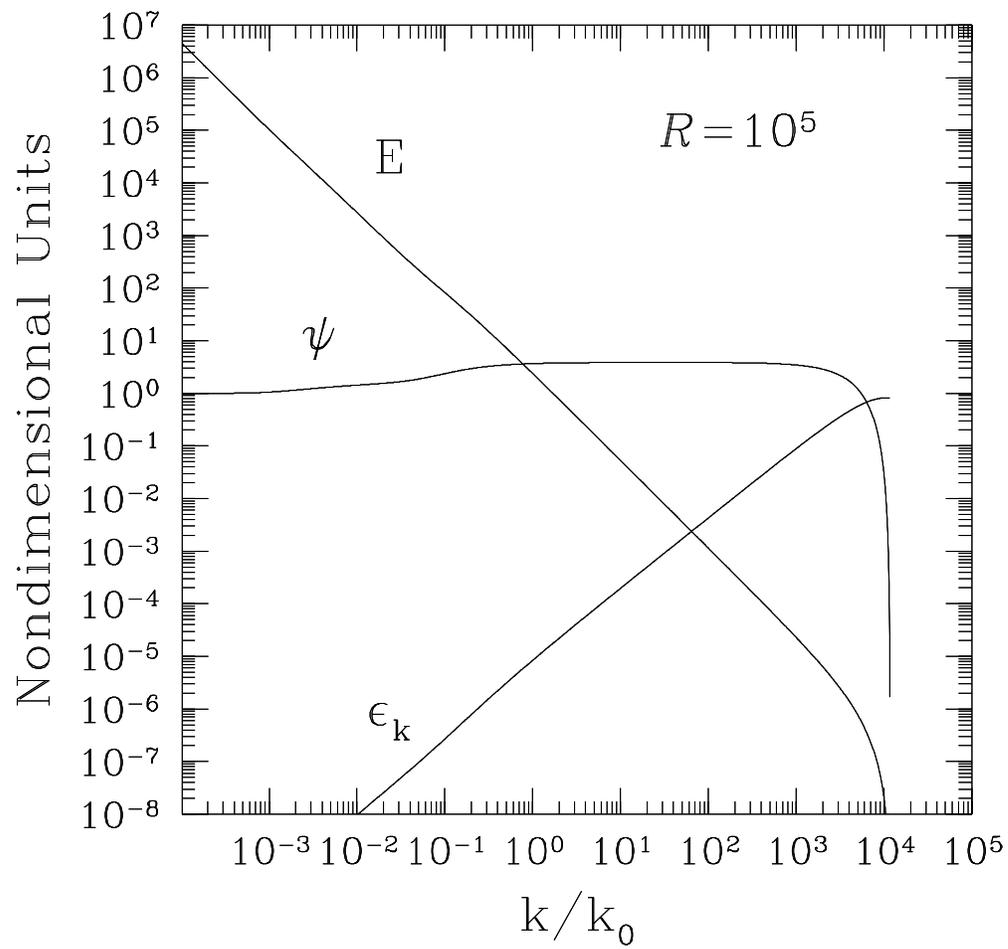
\begin{figure}
\caption{Solution of the steady-state dynamic equation, $\psi_k$, for 
Reynolds number $\R=10^5$ as a function of wavenumber, using
the grand source function shown in Fig. 1. The spectrum $E(k)=\psi_k^{2/3}k^{5/3}$ 
is also shown together with the dissipation rate $\epsilon_k$.}
\end{figure}

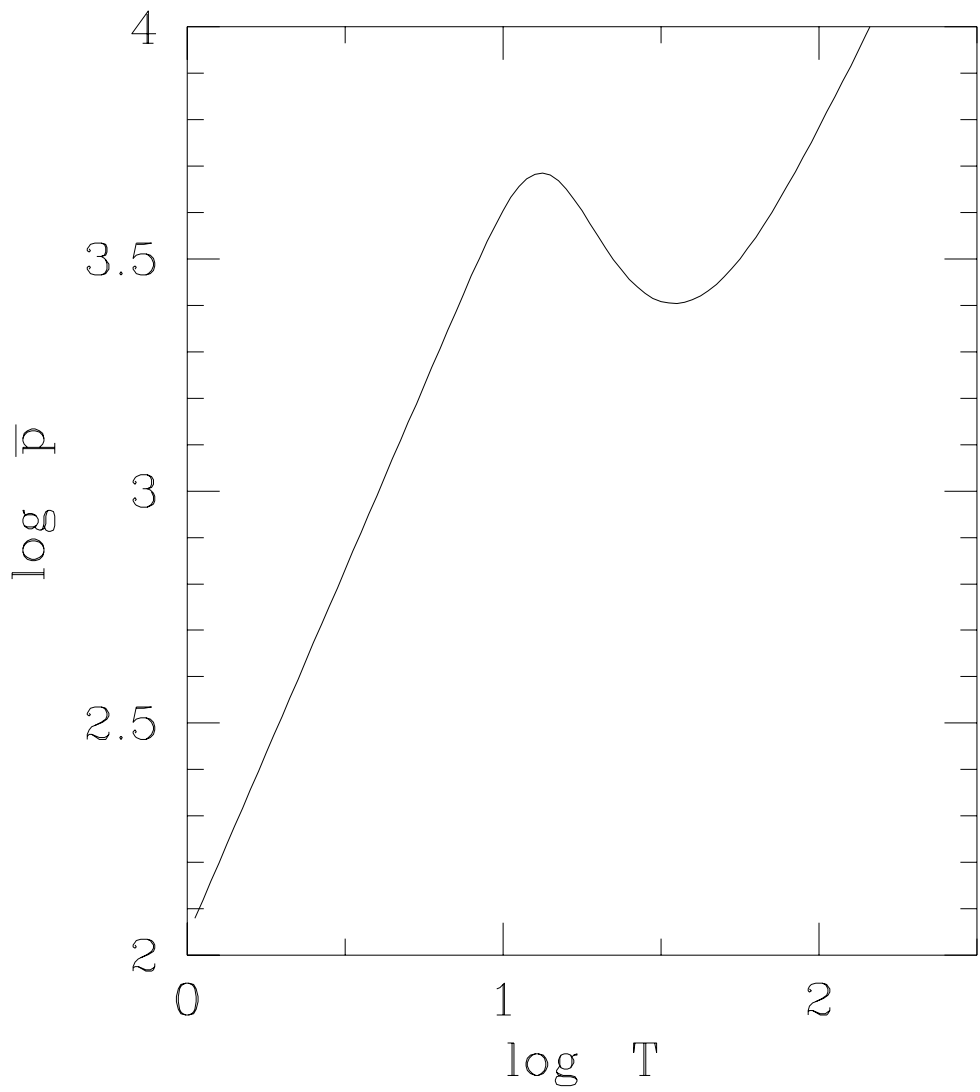
\begin{figure}
\caption{Pressure - temperature relation at $k=k_d$ for pumping number
$\S \sim 100$, Reynolds number $\R=10^5$, and eddy turnover time $t_e = 3\times
10^4$~yr.}
\end{figure}

\begin{figure}
\caption{Temperature as a function of wavenumber for pumping number
$\S \sim 100$, Reynolds number $\R=10^5$, and eddy turnover time $t_e = 3\times
10^4$~yr. The continuum structure is almost fractal in major portions of the   
diagram.}
\end{figure}

\begin{figure}
\caption{Density spectrum as a function of wavenumber for pumping number
$\S \sim 100$, Reynolds number $\R=10^5$, and eddy turnover time $t_e = 3\times
10^4$~yr.  In the range $-5 \leq log (k/k_d) \leq 0$, the approximate density 
power index is 0.55, again exhibiting approximate fractal behavior over a large
dynamic range.}
\end{figure}

\rotate[r]{
\vbox{
\begin{center}
\begin{tabular}{ccccc}
\multicolumn{5}{c}{\bf TABLE 1. SOURCE FUNCTIONS FOR TURBULENCE IN THE ISM
$^{*}$}\\
[12PT]
\hline
 & SUPERNOVAE & HII REGIONS & WINDS  & SUPERBUBBLES\\\hline\hline
Expansion Law &
$R = R_{SN}\left[ {4 t e \over 3 \tau_{SN}} - {1 \over
3}\right]^{3 \over 10}$ & $R = R_{HII}\left[ 1 + { 7 C_{II}t \over 4
R_{HII}}\right]^{4 \over 7}$ & $R = \left[ {L \over \rho}\right]^{ 1
\over 5} t^{ 3 \over 5}$ & $R =\left[ {125L_{SB} \over 154 \pi \rho}\right]^{ 1
\over 5} t^{ 3 \over 5}$ \\
&\\
 $R - \mu $ relation & $R = R_{0SN}\mu^{ -{ 2 \over 3}}$ & $R =
R_{0HII}\mu^{ -{ 4\over 3}}$ & $R = R_{0W}\mu^{ -{ 3 \over 2}}$ & $R =
R_{0SB}\mu^{ -{ 3 \over 2}}$ \\
& \\
 $R_{0i}$ & $R_{SN}\left[ 2 R_{SN} \over 5 c_s \tau_{SN}\right]^{ 3
\over 7}$& $\left[ {7 C_{II} \over 4C_{I}}\right]^{ 4 \over 3}R_S$ & $
\left[ 27 L \over 125 \rho c_s^3\right]^{1 \over 2}$ & $
\left[ 27 L_{SB} \over 154 \pi \rho c_s^3\right]^{1 \over 2}$ \\
& \\
 $\alpha_i$ & 9/7 & 4 & 9/2 & 9/2 \\
& \\
CANONICAL VALUES\\
 $\gamma_i$ (yr) & 1/30  & 1/60&1/30 & 1/3000
 \\
& \\
 $R_{0i}$ (pc)& 70 & 30 & 70 & 1000 \\
\hline
\end{tabular}
\end{center}
$^{*}$ See Appendix B for definition of various quantities
}}
\end{document}